\documentclass[twocolumn,superscriptaddress,english]{revtex4-1}
\usepackage{amsmath,amssymb}
\usepackage{graphicx}
\usepackage{subscript}
\begin{document}

%%%%%%%%%%%%%%%%%%%%%%%%%%%%%
%%% Title Page 
\title{Chimera states in a network-organized public goods game with destructive agents}

\author{Nikos E. Kouvaris} 
\affiliation{Departament de F\'isica de la Mat\`eria Condensada, Universitat de Barcelona, Mart\'i i Franqu\`es 1, 08028 Barcelona, Spain.}
\affiliation{Center for Brain and Cognition, Universitat Pompeu Fabra, Carrer de Tanger 122-140, 08018 Barcelona, Spain.}
\affiliation{Department of Information and Communication Technologies, Universitat Pompeu Fabra, Barcelona, Spain.}

\author{Rub\'en J. Requejo}
\affiliation{Departament de F\'isica de la Mat\`eria Condensada, Universitat de Barcelona, Mart\'i i Franqu\`es 1, 08028 Barcelona, Spain.}

\author{Johanne Hizanidis} 
\affiliation{Crete Center for Quantum Complexity and Nanotechnology, Physics Department, University of Crete, 71003 Heraklion, Greece.}
\affiliation{National Center of Scientific Research ``Demokritos'', 15310 Athens, Greece.}

\author{Albert D{\'i}az-Guilera} 
\affiliation{Departament de F\'isica de la Mat\`eria Condensada, Universitat de Barcelona, Mart\'i i Franqu\`es 1, 08028 Barcelona, Spain.}
\affiliation{Universitat de Barcelona Institute of Complex Systems (UBICS), Universitat de Barcelona, Barcelona, Spain.}

\date{\today}

%====== Abstract =====

\begin{abstract}
We found that a network-organized metapopulation of cooperators, defectors and destructive agents playing the public goods game with mutations, can collectively reach global synchronization or chimera states.
Global synchronization is accompanied by a collective periodic burst of cooperation, whereas chimera states reflect the tendency of the networked metapopulation to be fragmented in clusters of synchronous and incoherent bursts of cooperation. 
Numerical simulations have shown that the system's dynamics alternates between these two steady states through a first order transition. 
Depending on the parameters determining the dynamical and topological properties, chimera states with different numbers of coherent and incoherent clusters are observed. 
Our results present the first systematic study of chimera states and their characterization in the context of evolutionary game theory.
This provides a valuable insight into the details of their occurrence, extending the relevance of such states to natural and social systems. 
\end{abstract}

\maketitle

\section{Introduction}
The public goods game (PGG) provides a classical example that describes the evolutionary dynamics of competing species or strategies in biological and social systems \cite{Hofbauer1998,sigmund:2010}. 
Usually this game is played by {\it cooperators}, which create public goods at a cost to themselves, and {\it defectors}, which enjoy the benefits but do not pay any cost \cite{sigmund:2010}. 
Then cooperation extinguishes and public goods creation vanishes in the so called {\it tragedy of the commons} \cite{hardin:1968}.  
However, the inclusion of a third non-participating strategy allows for a sequential dominance of cooperation, defection and abstention from the game \cite{hauert:2002a, hauert:2002b, arenas:2011, requejo:2012c}. 
This latter behavior resembles the rock-paper-scissors game \cite{Hofbauer1998} which has been found experimentally in the three competing strains of E. coli \cite{kerr2002} as well as in social groups with cooperators, defectors and volunteers \cite{Semmann2003}.

It has been shown that mutations among strategies could give rise to more complex dynamical behavior, like the emergence of self-sustained oscillations via a supercritical Hopf bifurcation \cite{mobilia2010, arenas:2011, requejo:2012c, toupo2015}. Moreover, spontaneous formation of complex patterns has been studied in spatially extended ecological systems \cite{vickers1993,wakano:2009,wakano:2011}. 
Non-trivial spatiotemporal patterns of synchronized action and their evolutionary role were also reported \cite{szabo:2016}.
Nevertheless, other aspects of complexity and the emergence of self-organization by means of synchronization \cite{arenas:2008} and chimera states \cite{panaggio:2015} have not been investigated intensively in the context of evolutionary game theory. 
Our study contributes to the acquisition of new findings towards this direction.

Chimera states are characterized by the coexistence of coherent and incoherent behavior in systems of coupled oscillators. 
They were initially reported for identical phase oscillators \cite{kuramoto:2002}, where the nonlocal coupling was thought to be the source of this counter-intuitive phenomenon~\cite{abrams:2004}. 
However, they have been recently found in systems with global \cite{schmidt:2014,sethia:2014,yeldesbay:2014,boehm:2015} and purely local coupling \cite{laing:2015,hizanidis:2016a,clerc:2016}.
Although, most works on chimera states consider simple network topologies (see \cite{panaggio:2015} and references within), recently, they have been found in real networks, like the {\it C.Elegans} neural connectome \cite{hizanidis:2015a, hizanidis:2016} and the cat cerebral cortex~\cite{santos:2016}.
It has been suggested that chimera states may be related to bump states in neural systems \cite{laing:2001,sakaguchi:2006}, the phenomenon of unihemispheric sleep~\cite{rattenborg:2000}, or epileptic seizures~\cite{andrzejak:2016}.
For finite systems chimera states are known to be chaotic transients~\cite{wolfrum:2011}, which can be stabilized by various recently
developed control schemes \cite{sieber:2014,bick:2015,isele:2015,omelchenko:2016}. 
The existence of chimera states has also been verified experimentally over the last years in various settings \cite{hagerstrom:2012,martens:2013,tinsley:2012,nkomo:2013,schmidt:2014,wickramasinghe:2013}. 

Here we study the emergence of collective phenomena, and specifically chimera states, in a PGG with mutations \cite{arenas:2011} which is organized on a ring network with nonlocal connections. 
In each node of the network-organized PGG the species can select among different strategies as determined by the replicator equation \cite{Hofbauer1998,sigmund:2010}. 
They are allowed also to mutate into one another with a uniform mutation rate.
Moreover, the network connectivity structure defines a mutual influence among strategies across the network nodes. 
The latter process, under appropriate conditions \cite{wakano:2009,requejo:2015}, resembles the diffusion of species across the network.
We show that the considered system exhibits synchronization and chimera states, and promotes, respectively, bursting oscillations of cooperation either globally or in regions separated by incoherent clusters.

\section{Replicator-mutator dynamics: Robust evolutionary cycles and self-sustained oscillations}
We assume a large well-mixed population of cooperators, defectors and destructive agents whose interactions are governed by a PGG \cite{arenas:2011}. 
\begin{figure}[b!]
\includegraphics{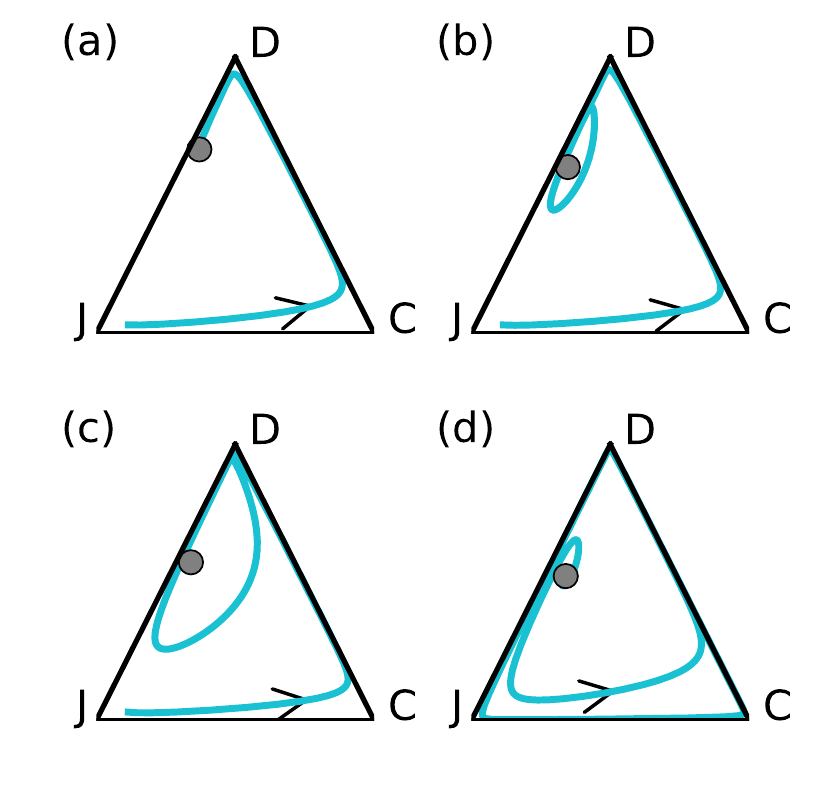} 
\caption{The phase space of the replicator-mutator dynamics, 
Eqs. \eqref{eq:localdyn}, exhibits various attractors which correspond to
(a) a stable focus; $d=0.25$ and $\mu=0.006$, 
(b) a limit cycle; $d=0.27$ and $\mu=0.004$,
(c) a limit cycle; $d=0.3$ and $\mu=0.003$,
(d) a limit cycle approaching a heteroclinic orbit; $d=0.4$ and $\mu=0.001$. 
Trajectories are projected into a simplex whose corners correspond to the dominance of cooperators (C), defectors (D) or destructive agents (J). 
Other parameters are $n=5$ and $r=3$.}
\label{fig:phasespace}
\end{figure} 
At each round of the game a group of $n$ individuals is randomly sampled: Cooperators from this group pay a cost $c$ and create a benefit $b=rc$ (with $r>1$) which is distributed equally among all participants of the group. 
Defectors receive their share from the benefits without paying any cost.
Destructive agents, without receiving any benefits, induce a damage $d$ into the game which is shared equally by cooperators and defectors.
The fitnesses of the individuals in a PGG determine their evolutionary fate, and are calculated as the average payoff of each strategy after its participation in many interaction groups, which for large populations $N\gg1$ (c.f. \cite{arenas:2011}) results in,
\begin{subequations}\label{eq:payoffs}
\begin{eqnarray}
P_x &=&
    r \frac{x}{1-z} \left[ 1 - \frac{1-z^n}{n(1-z)} \right] 
    + \frac{r}{n} \frac{1-z^n}{1-z} - 1 \nonumber\\
    &&-\: d \left( \frac{1-z^n}{1-z} - 1\right)\,,\label{eq:payoffA}\\
P_y &=& P_x + 1 - \frac{r}{n} \frac{1-z^n}{1-z}\,,\label{eq:payoffB}\\
P_z &=& 0 \,,\label{eq:payoffC}
\end{eqnarray}
\end{subequations}
\begin{figure}[b!]   
\includegraphics{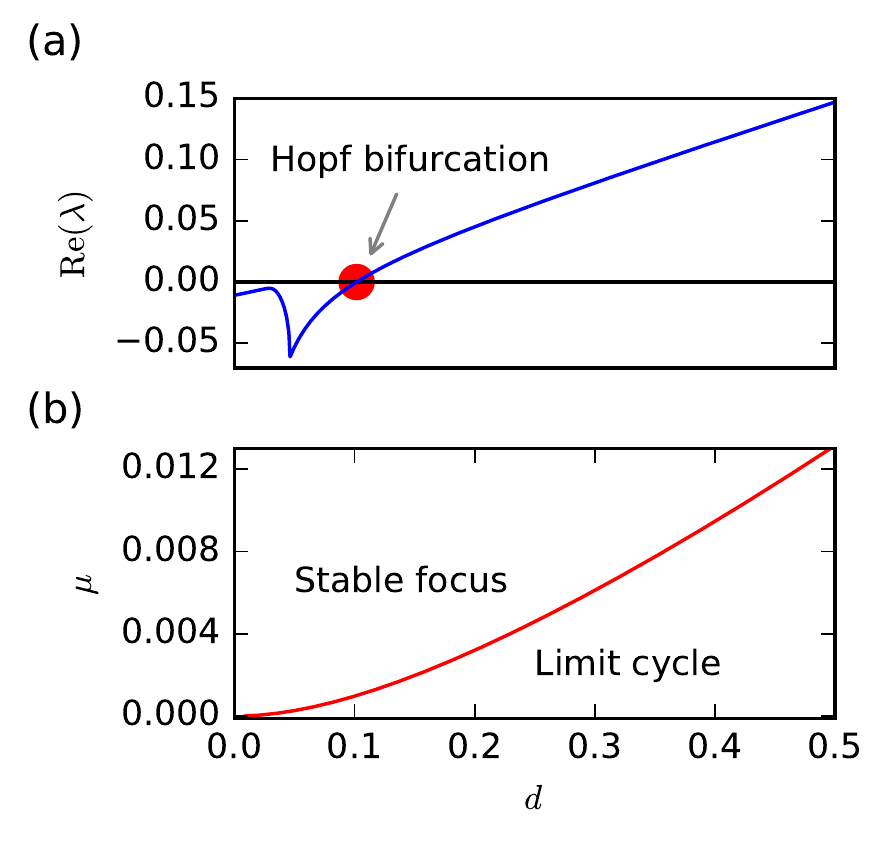}
\caption{Stability analysis of the system \eqref{eq:localdyn}.
(a) The fixed point of the system has complex conjugate eigenvalues whose real part is shown as a function of the damage parameter $d$ for the mutation rate $\mu=0.001$. 
A stable focus loses its stability via a supercritical Hopf bifurcation (red dot) and becomes unstable giving rise to a limit cycle.
(b) Continuation of the Hopf point determines the curve which separates different dynamical regimes in the parameter space $d$--$\mu$. 
Other parameters are $n=5$ and $r=3$.}
\label{fig:hopf} 
\end{figure} 
\noindent where $x$, $y$ and $z$ are the fractions of cooperators, defectors and destructive agents (or the relative frequencies of individuals playing each strategy), respectively; $n$ is the group size and $d$ is the total damage that destructive agents inflict to the participants of the game. 
Without loss of generality, we set the cost paid by the cooperators to unity, $c=1$. 
As a consequence, the multiplicative factor $r$ now represents the benefit produced per cooperator in the group. 

The evolution of the three strategies can be studied by the replicator-mutation dynamics \cite{taylor:1978, hofbauer:1979} given by,
\begin{subequations}\label{eq:localdyn}
\begin{eqnarray}
\dot{x} &=&
	x (P_x -\bar{P}) 
	+ \mu (1 - 3 x)
	\,,\label{eq:localdynA}\\
\dot{y} &=&
	y (P_y -\bar{P}) 
	+ \mu (1 - 3 y) 
	\,,\label{eq:localdynB}\\
\dot{z} &=& 
	z (P_z -\bar{P}) 
	+ \mu (1 - 3 z) 
	\,,\label{eq:localdynC}
\end{eqnarray}
\end{subequations}
\noindent where $\bar{P} = x P_x + y P_y + z P_z$ is the average payoff of the population at a given time. 
Obviously $x+y+z=1$; this allows to reduce the dimensionality of the phase space and analyze the dynamics of three strategies only by investigating $x$ and $y$. \
In each equation \eqref{eq:localdyn}, in addition to the replication term which accounts for the variation of the fractions of individuals due to the selection process (first term on the right hand side), mutations are also included (second term) and represent random changes between the strategies at a rate $\mu$. 
This system has one non-trivial and three trivial fixed points (see figure \ref{fig:phasespace}). 
The trivial fixed points are saddles and represent the dominance of cooperators ($C;\ x=1$), defectors ($D;\ y=1$) or destructive agents ($J;\ z=1$). 
The non-trivial point (gray dot) can behave as a stable focus (see e.g. figure \ref{fig:phasespace}(a)) that attracts all the trajectories or as an unstable focus  (see e.g. figure \ref{fig:phasespace}(b)--(d)) that repels the trajectories, which however, are confined within the heteroclinic cycle, hence they are attracted to a stable limit cycle.

Linear stability analysis has shown that a supercritical Hopf bifurcation occurs for increasing $d$ or decreasing $\mu$, beyond which self-sustained oscillations spontaneously emerge. 
Figure \ref{fig:hopf}(a) shows the Hopf bifurcation point (red dot) for a fixed mutation rate, while the continuation of the Hopf point determines the curve which separates different dynamical regimes in the parameter space $d$--$\mu$ (see figure \ref{fig:hopf}(b)). 
The amplitude and the period of the limit cycles becomes larger as the parameters $d$ and $\mu$ lie further from the Hopf point.

\section{Network-organized replicator-mutator dynamics}
Here we consider a metapopulation of individuals which are organized on ring networks with nonlocal 
connections \cite{omelchenko:2013,hizanidis:2015}. 
Each node of such networks is occupied by a large well-mixed population of individuals which interact internally according to a PGG as described above. 
In addition to the local interactions ---that is, replications and mutations--- the populations in each node take into account 
the strategies followed by the populations in their connected nodes. 
In the ring networks considered here, the population size in the nodes is assumed to be constant. 
Therefore, the overall process can be described by the following equations:
\begin{figure*}[t!]    
\includegraphics{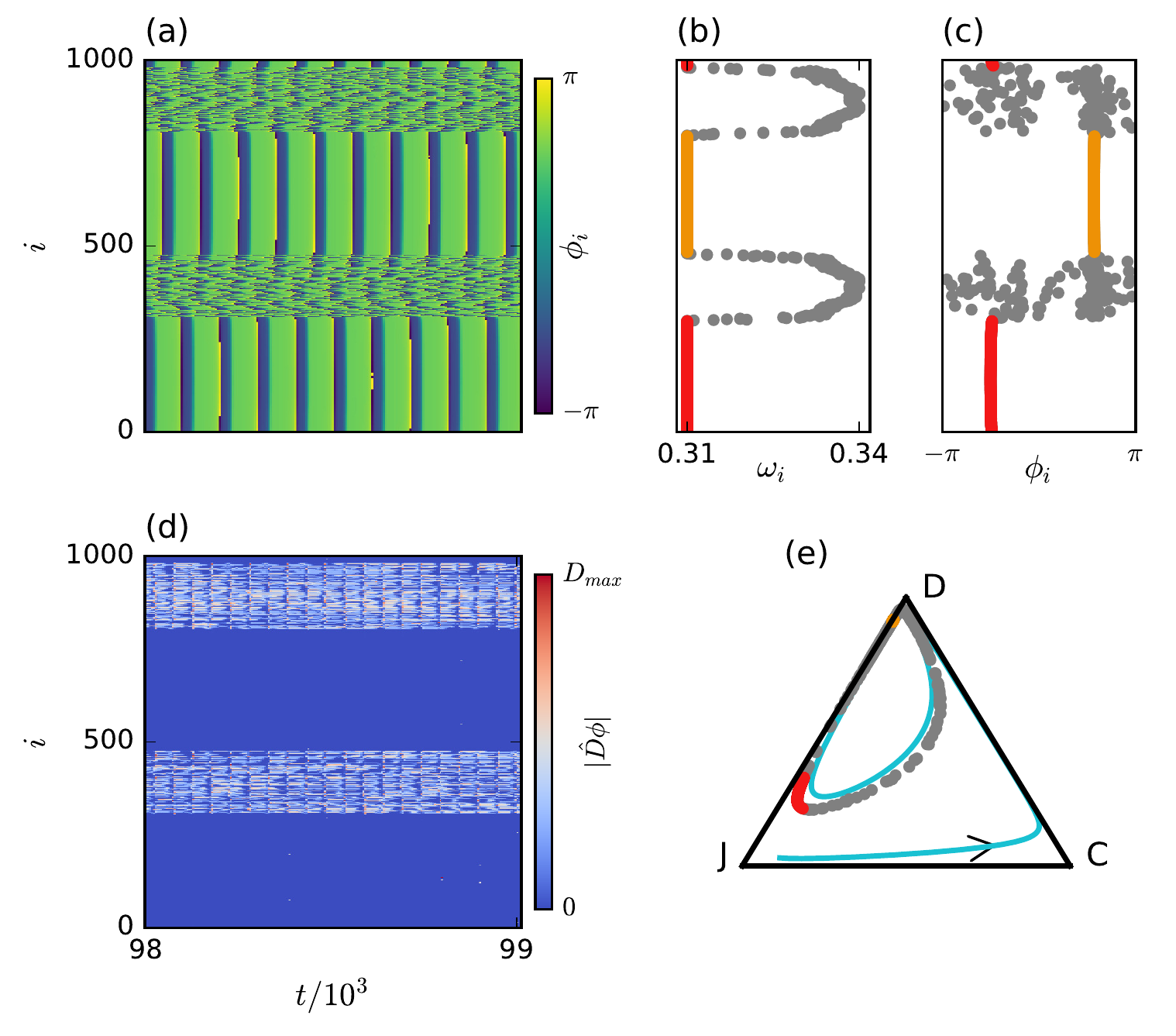}
\caption{(Color online) Clustered chimera state with two (in)coherent clusters. (a) Space-time plot of the phase $\phi$, (b) $\omega$-profile, (c) chimera snapshot, (d) space-time plot of spatial coherent index $|\hat{D}\phi(t)|$, and (e) phase space
representation of a chimera state. 
Other parameters are $N=1000$, $R=320$, $\sigma=0.008$, $\mu=0.001$, $d=0.23$, $n=5$, and $r=3$. 
(See also Supporting Movie S1)}
\label{fig:multiplot}
\end{figure*}
\begin{subequations}\label{eq:ringdyn}
\begin{eqnarray}
\dot{x}_i &=&
	x_i (P_{x,i} -\bar{P}_i) 
	+ \mu (1 - 3 x_i)
	+ \frac{\sigma}{2R} \sum_{j=i-R}^{j=i+R}\! (x_j - x_i)\,,
	\label{eq:ringdynA}\\
\dot{y}_i &=& 
	y_i (P_{y,i} -\bar{P}_i) 
	+ \mu (1 - 3 y_i) 
	+ \frac{\sigma}{2R} \sum_{j=i-R}^{j=i+R}\! (y_j - y_i)\,,
	\label{eq:ringdynB}\\
z_i &=& 1 - x_i - y_i \,,\label{eq:ringdynC}
\end{eqnarray}
\end{subequations}
\noindent where the summation terms account for the mutual influence of strategies between populations in connected nodes and $\sigma$ characterizes the strength of this influence.
Taking into account Eq.~\eqref{eq:ringdynC} the latter process is equivalent to the diffusion of cooperators and defectors across the network (c.f. \cite{wakano:2009,requejo:2015}).

In general, an increasing coupling strength $\sigma$ in the system \eqref{eq:ringdyn} results in synchronization of the metapopulation, where the fractions of cooperators, defectors and destructive agents in each node oscillate with the same phase and amplitude. 
However, the nonlocal topology of the ring network can induce non-trivial collective phenomena like chimera states. 
In the following, we focus on the analysis of these states.

As a measure indicating the existence of a chimera state we employ the {\it mean phase velocity}
of each oscillator~\cite{kuramoto:2002,omelchenko:2013}: 
\begin{equation}
\omega_i = \frac{2 \pi M_i}{\Delta T}\,,
\label{eq:omega}
\end{equation}
\noindent where $M_i$ is the number of periods of the $i$-th oscillator during a time interval $\Delta T$. 
The typical profile of $\omega_i$ in the case of 
a chimera state is flat in the synchronous domains and arc-shaped in the incoherent ones. 
In addition to the mean phase velocity, we calculate the classification measures for chimera states developed recently by Kemeth {\it et al.} in ~\cite{kemeth:2016}.
In particular, we employ the local curvature of the phases of the oscillators as a measure for the spatial coherence. 
The phase of each oscillator is defined as,
\begin{equation}
\phi(t) = \arctan\left(\frac{y(t)-\langle y\rangle_t}{x(t)-\langle x\rangle_t}\right)\,,
\label{eq:phase}
\end{equation}
\noindent where $\langle x\rangle_t,\ \langle y\rangle_t$ denote time averages.
In the ring networks considered here, we calculate the local curvature at each node $i$ by applying the discrete Laplacian operator $\hat{D}$ on each snapshot $\{\phi_1,\phi_2,\ldots,\phi_N\}$ at time $t$. 
This operator reads:
\begin{equation}
\hat{D}\phi(t) := \lbrace \phi_{i-1}(t) - 2 \phi_i(t) + \phi_{i+1}(t) \,,\,  \forall \, i \in (1,N)\, \rbrace \,,
\label{eq:loc_curv}
\end{equation}
\noindent where $\phi(t)$ denotes the spatial distribution of the phases in one spatial dimension with periodic boundary conditions at time $t$. 
For the nodes in the synchronous/coherent clusters $\phi_\text{coh}$ it holds that $|\hat{D}\phi_\text{coh}(t)|=0$, while for the nodes in the incoherent clusters $\phi_\text{incoh}$, $|\hat{D}\phi_\text{incoh}(t)|$ is finite and has pronounced fluctuations. 
The maximum value $D_{max}$ of $|\hat{D}\phi(t)|$ corresponds to the local curvature of nodes whose two nearest neighbors have the maximum phase difference. 

\begin{figure*}[t!]  
\center  
\includegraphics{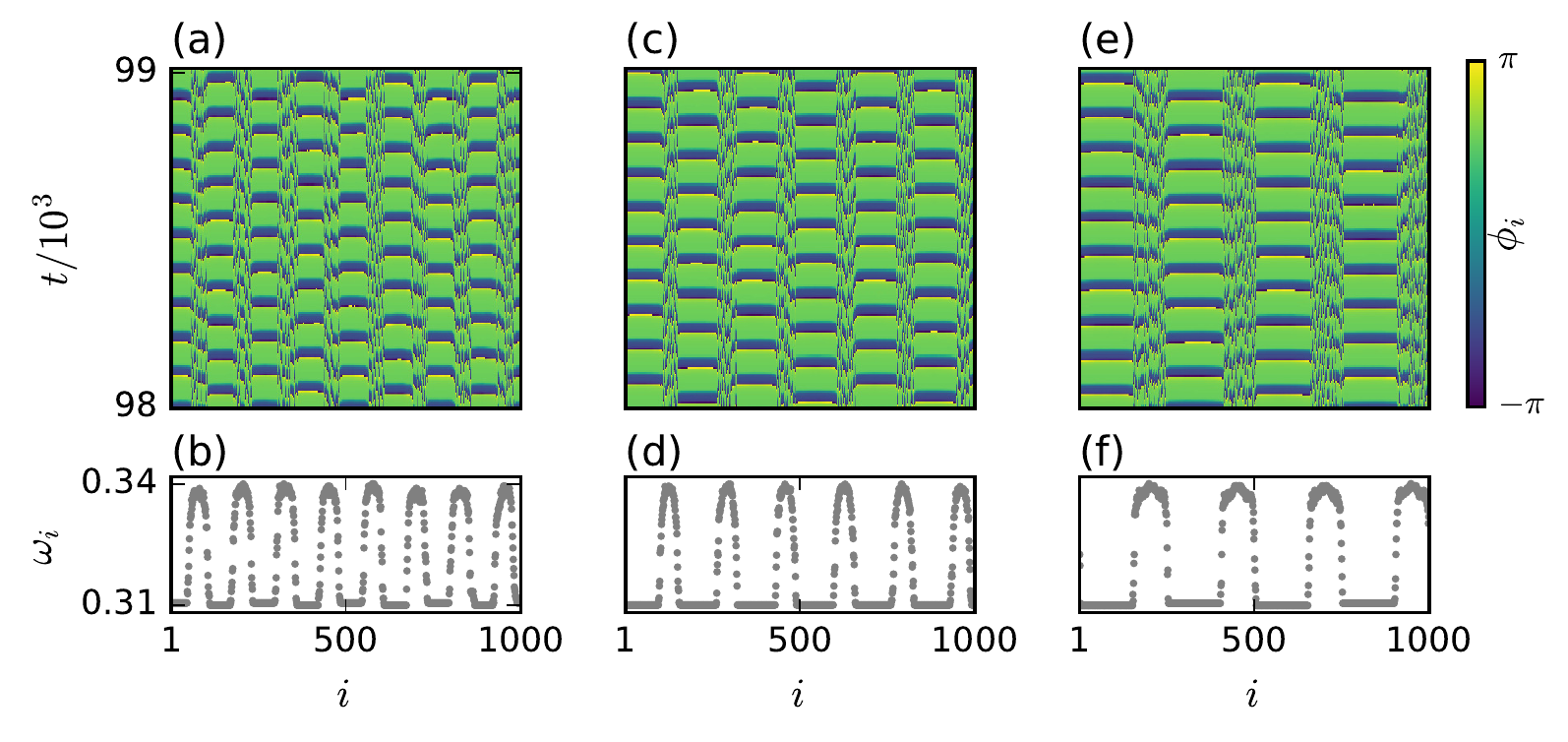}
\caption{(Color online) Multi-clustered chimera states are shown for different values of $R$. 
(a) Space-time plot of the phase $\phi$ for $R=70$ and (b) corresponding $\omega_i$
profile; this chimera state has eight clusters of (in)coherent nodes. 
Similar plots are shown in (c)--(d) for $R=110$, and in (e)--(f) for $R=150$, where the corresponding chimera states have six and four clusters of incoherent nodes, respectively. 
Other parameters are $N=1000$, $\sigma=0.008$, $\mu=0.001$, $d=0.23$, $n=5$, and $r=3$.}
\label{fig:spacetime}
\end{figure*}

The local curvature defined above allows for a clear representation and characterization of the obtained chimera states. 
Figure~\ref{fig:multiplot} shows a typical chimera state emerging from the dynamics of our model:
In (a) we see the space-time plot of the phase $\phi$, while (b) and (c) show the corresponding mean phase velocity profile and a snapshot at a given time instance. 
Figure~\ref{fig:multiplot}(d) shows the space-time evolution of the spatial coherence index (Eq.~\ref{eq:loc_curv}) and figure~\ref{fig:multiplot}(e) illustrates a single time snapshot of the chimera state in the phase space. 
The gray dots correspond to the incoherent cluster, the red and orange segments refer to the coherent domains, and the solid line marks the orbit of the uncoupled unit.

In the example of figure~\ref{fig:multiplot}, the observed chimera state has two (in)coherent regions. 
The multiplicity of a chimera state (number of synchronous clusters) may be manipulated by varying the coupling range of each node. 
This results in the formation of multi-clustered (or multi-headed) chimeras reported in many systems~\cite{omelchenko:2013,vuellings:2014,maistrenko:2014}. 
The effect of the coupling range is illustrated in figure~\ref{fig:spacetime}, where the space-time plots for phase $\phi$ and the corresponding mean phase velocity profiles are shown for three different values of $R$.
Note that the coherent regions are always in antiphase~\cite{sethia:2008}, which explains also the even number of (in)coherent clusters in the obtained chimeras.

Based on the local curvature we can measure the relative size of the spatially coherent (i.e. synchronized) clusters at each time step. 
For this purpose we consider the normalized probability function $g$ of $|\hat{D}\phi(t)|$, $g(|\hat{D}\phi(t)|=0)$; it equals $0$ in a non-synchronous system and $1$ in a fully synchronized one. 
Any value of $g(|\hat{D}\phi(t)|=0)$ between $0$ and $1$ indicates coexistence of coherence and incoherence, i.e. a chimera state. 
The definition of spatial coherence or incoherence is not absolute, but depends on the maximum curvature of the system. 
Therefore this index is defined with the threshold $\delta = 0.01 D_{max}$ as: 
\begin{equation}
g_0 := \sum_{|\hat{D}\phi(t)|=0}^{\delta}\! g(|\hat{D}\phi(t)|) \,.
\label{eq:g0}
\end{equation}
Apart from the spatial coherence, we also calculate the temporal coherence as an indication
for a chimera state, based on the pairwise correlation coefficients~\cite{kemeth:2016}:
\begin{equation}
\rho_{ij} = \frac{\langle (\Phi_i - \langle\Phi_i\rangle) (\Phi_j - \langle\Phi_j\rangle) \rangle}{(\langle\Phi_i^2\rangle-\langle\Phi_i\rangle^2)^{1/2} (\langle\Phi_j^2\rangle-\langle\Phi_j\rangle^2)^{1/2}} \,,
\label{eq:corrcoef}
\end{equation}
\noindent where $\Phi_i$, $\Phi_j$ are the time series of the phases of two oscillators in the nodes $i$ and $j$, respectively. 
The normalized distribution function $h(|\rho|)$ is a measure for the correlation in time and the  percentage of the time-correlated oscillators is given by:
\begin{equation}
h_0 := \left( \sum_{|\rho|=\gamma}^{1}\! h(|\rho|) \right)^{1/2}\,,
\label{eq:h0}
\end{equation}
\noindent where the coherent accuracy for correlated oscillators is $\gamma=0.99$.

\begin{figure}[th!]   
\center 
\includegraphics{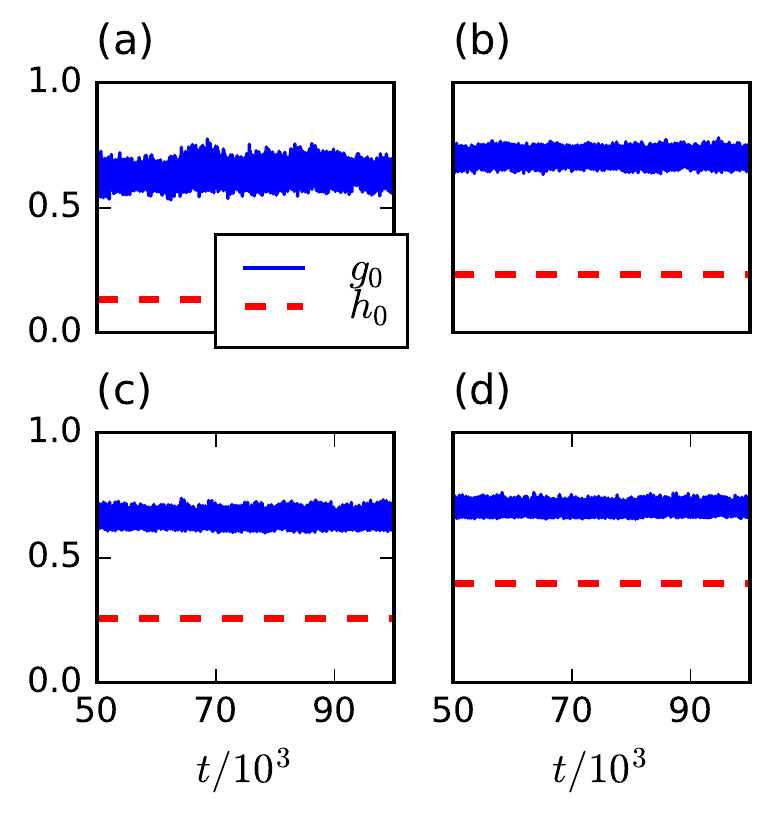}
\caption{Measures for spatial ($g_0$) and temporal ($h_0$) coherence for the chimera states shown in figure \ref{fig:multiplot} and figure \ref{fig:spacetime} for (a) $R=70$, (b) $R=110$, (c) $R=150$, (d) $R=320$. 
Other parameters are $N=1000$, $\sigma=0.008$, $\mu=0.001$, $d=0.23$, $n=5$, and $r=3$.}
\label{fig:g0h0}
\end{figure}

The influence of the coupling range on the spatial and temporal coherence of the observed dynamics is depicted in figure~\ref{fig:g0h0}.
Both measures, $g_0$ and $h_0$, are within the parameter range that ensures the existence of chimera states. 
As $R$ increases, so does the
size of the coherent clusters, which is reflected by the increasing values of $h_0$ and $g_0$.
Moreover, in all cases $h_0$ is fixed in time and $g_0$ fluctuates slightly around a constant value (this effect diminishes for larger $R$);
therefore, the chimera states are {\it stationary} and {\it static} according to the classification scheme of \cite{kemeth:2016}.

\section{Abrupt transitions between chimera states and synchronization}
\begin{figure*}[t!]    
\includegraphics{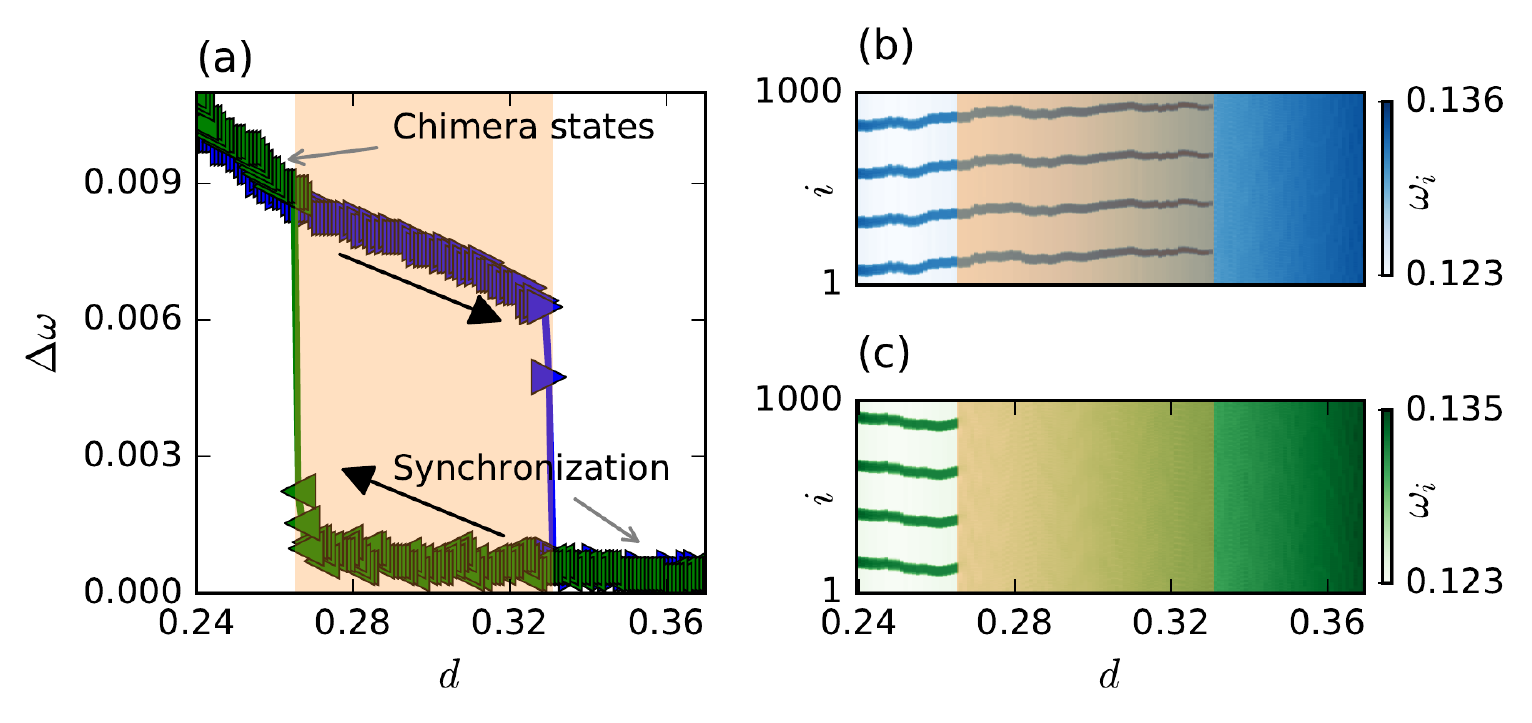}
\caption{(Color online) 
Increasing (blue color) and decreasing (green color) values of $d$ cause first order transitions between chimera states and synchronization. 
This reveals a hysteresis loop, shown in (a) for a ring of $N=1000$ nodes and coupling range $R=180$. 
For values of $d$ within the orange colored region the system \eqref{eq:ringdyn} can exhibit either chimera states with four (in)coherent clusters or synchronization.
Mean phase velocity $\omega_i$ of each population $i$ is depicted as a function of $d$ for both scenarios of (b) increasing and (c) decreasing values. 
In all plots the orange colored region corresponds to the same interval of values of $d$.
Other parameters are $\sigma=0.008$, $\mu=0.001$, $n=5$, and $r=3$.}
\label{fig:hysteresis_d}
\end{figure*}
The above analysis elucidates that the replicator-mutator dynamics of the PGG organized on ring networks with nonlocal coupling support either synchronization or chimera states, whose features depend on parameters determining dynamical and topological properties. 

In the following, a detailed analysis of this dependence will be presented by focusing on two parameters, the damage $d$ and the coupling range $R$.
For our analysis we take into account that the populations in the nodes of coherent and incoherent domains oscillate with mean phase velocities $\omega_{\rm coh}$ and $\omega_{\rm incoh}$, respectively. 
The faster populations in the incoherent domain oscillate with $\omega_{\rm incoh}^{\rm max}$.
Therefore, by looking at the difference 
\begin{equation}
\Delta\omega = \omega_{\rm incoh}^{\rm max} - \omega_{\rm coh}
\label{eq:delta_omega}
\end{equation}
\noindent one can ensure that chimera states exist when $\Delta\omega$ is larger than a certain threshold.

Extensive numerical simulations have revealed that a small change in the parameter $d$ can cause suddenly an abrupt, first order transition between synchronized and chimera states, which is characterized by a hysteresis loop (see Figure \ref{fig:hysteresis_d}(a) orange colored area). 

Starting from an initial configuration of a chimera state with four (in)coherent clusters we perform numerical simulations (continuation) by increasing and then decreasing slowly the damage $d$ for fixed coupling range $R=180$. 
Figure \ref{fig:hysteresis_d}(b) shows that a gradual increase of the damage (which shifts the system further from the Hopf bifurcation) changes slightly the position and the size of the incoherent clusters up to a critical value for which an abrupt transition occurs suddenly and brings the system to a synchronized state where it remains thereafter.
Figure \ref{fig:hysteresis_d}(c) shows an opposite (but qualitatively similar) scenario: Decreasing the damage of the game gives rise to an abrupt transition which brings the system back to a chimera state with four (in)coherent clusters. 
However, this second transition takes place at a different value of $d$, resulting in the observed hysteresis loop (c.f., \cite{vuellings:2014}).

\begin{figure*}[t!]    
\includegraphics{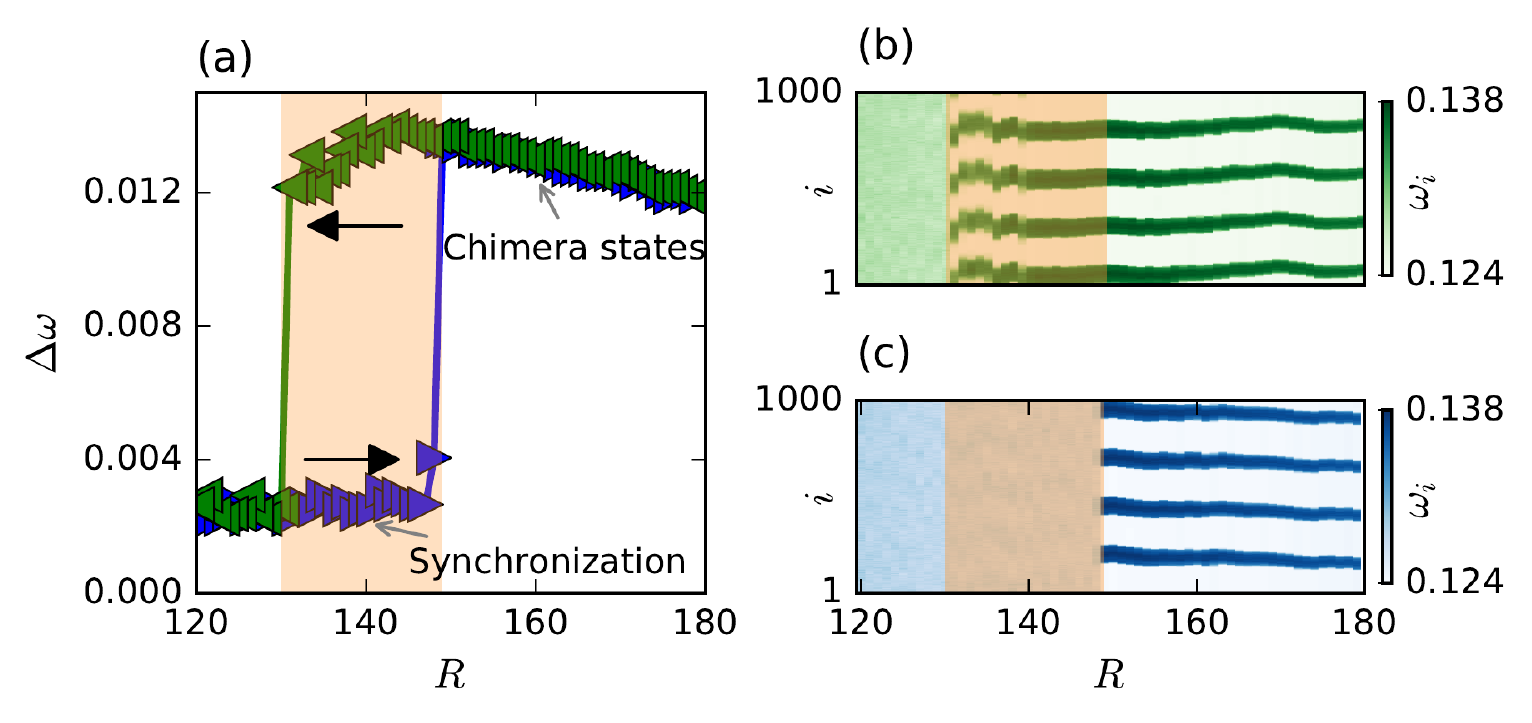}
\caption{
(Color online)
Decreasing (green color) and increasing (blue color) values of $R$ cause first order transitions between chimera states and synchronization. 
This reveals a hysteresis loop, shown in (a) for a ring of $N=1000$ nodes and $d=0.23$.
For values of $R$ within the orange colored region the system \eqref{eq:ringdyn} can exhibit either chimera states with four (in)coherent clusters or synchronization.
The mean phase velocity profile $\omega_i$ of each population $i$ is depicted as a function of $R$ for both scenarios of (b) decreasing and (c) increasing values.
In all plots the orange colored region denotes the same interval of $R$ values.
Other parameters are $\sigma=0.008$, $\mu=0.001$, $n=5$, and $r=3$.}
\label{fig:hysteresis_R}
\end{figure*}

Starting from the same initial configuration as above, we now perform numerical continuation by decreasing and then increasing the coupling range $R$ for fixed damage $d=0.23$. 
Figure \ref{fig:hysteresis_R}(a) shows that an abrupt transition from a chimera to a  synchronized state and back occurs suddenly and is characterized by a hysteresis loop. 
Like in the case of varying $d$, there is a window of values for the coupling range (orange colored area) where for the same topology (i.e. same $R$) the system can either be self-organized into a chimera state with four (in)coherent clusters or be synchronized, depending on the initial conditions.
Figures \ref{fig:hysteresis_R}(b) and (c) illustrate the mean phase velocity $\omega_i$ of each population $i$ as a function of $R$. 
This allows to discriminate the existence of (in)coherent clusters (i.e. existence of chimera states), their position and their size, for both directions of the continuation.

Numerical continuation between different limits for $d$ or $R$ has revealed that, in general, different initial configurations give rise to various transitions between synchronization and chimera states. 
Interestingly, transitions between chimera states with different number of (in)coherent clusters were also found (see Supporting Information).

\section{Discussion}
For the first time we report on the existence of synchronization and chimera states in ring networks with nonlocal coupling obeying the replicator-mutator dynamics of a PGG with cooperators, defectors and destructive agents. 
Our findings reflect the tendency of metapopulations to evolve collectively in a coherent way or be fragmented in clusters of synchronous and incoherent behavior.
The transition between these steady states occurs through an abrupt first order transition.

A systematic numerical analysis has revealed that chimera states are stationary and static, while the number of (in)coherent clusters varies depending on the coupling range $R$, and on the parameters that determine the local dynamics. 
Interestingly, the first order transitions which shift the system between steady states are characterized by strong hysteresis loops, where multistability is observed. 
In the hysteresis loop, depending on the initial conditions, either global synchronization or chimeras with varying number of (in)coherent clusters are achieved.

Our study provides for a new framework for the analysis of spontaneously emergent spatiotemporal phenomena in game theory, and particularly their effect on the cooperation-defection-destruction cyclic dynamics triggered by damaging individuals. 
Since synchronized or incoherent actions can influence cooperation and the efficiency of groups \cite{Wiltermuth:2009}, the appearance of the chimera states, in which the cyclic dynamics is accelerated, may have a relevant impact on such public goods creation and, hence, on the speed of evolution and innovation. 
Therefore, the stylized model presented here, may be adapted and completed to find applications in biological, social or economic systems.
As an example, the results found here can support the design of feedback schemes which, by promoting modifications in the strategy (dynamics) or in the connectivity structure (topology), control the collective --global or clustered-- behavior of metapopulations in order to, for instance, diminish long destructive periods or enhance innovation, as well as on biological synthetic systems, where chimera states may speed up reaction processess and evolution.

\section{Acknowledgments}
N.E.K., R.J.R. and A.D.-G. acknowledge financial support by the LASAGNE (Contract No.318132) EU-FP7 project.
N.E.K. and A.D.-G. also acknowledge financial support by the  MULTIPLEX (Contract No.317532) EU-FP7 project, the MINECO (projects FIS2012-38266 and FIS2015-71582), and the Generalitat de Catalunya (project 2014SGR-608).
J.H. acknowledge financial support by the SIEMENS research program on ``Establishing a Multidisciplinary and Effective Innovation and Entrepreneurship Hub''.

\end{document}